# En busca de los efectos gravitacionales de Birkhoff: La ultracentrífuga Transelástica.

S. Galindo*, Carlos Cabrera♣ y Jorge L. Cervantes-Cota*

*Departamento de Física y ♣Departamento de Sistemas Electrónicos
Instituto Nacional de Investigaciones Nucleares,
Km. 36.5 Carretera México- Toluca 52045, México,
e-mail : salvador.galindo@inin.gob.mx

Este artículo examina el origen, desarrollo, culminación y desenlace del proyecto de la ultracentrífuga transelástica del Centro Nuclear de México ocurridos entre 1971 y 1986. El proyecto tuvo su origen en la búsqueda de un efecto que supuestamente comprobaría la teoría de gravitación de Birkhoff sobre la Relatividad General de Einstein. Para este propósito se diseñó y construyó una extraordinaria ultracentrífuga que mereció el Premio Nacional de Instrumentación (México) 1973. La ultracentrífuga también fue usada para investigar la factibilidad de enriquecer uranio por centrifugación en estado sólido. Se obtuvo uranio altamente enriquecido, aunque en pequeñas cantidades.

This work reviews the origin, development, completion, and outcome of a trans-elastic ultracentrifuge project of Mexico´s Nuclear Center through 1971 to 1986. The project had its origin in the search for an effect that supposedly would validate Birkhoff´s gravity theory over Einstein´s General Relativity. For this purpose an extraordinary ultracentrifuge was built which deserved the 1973 National Award for Instruments (Mexico). The ultracentrifuge was also used to investigate the feasibility of uranium enrichment by solid state centrifugation. Highly enriched uranium was obtained, but in small quantities.

1. **Introducción.**

La ejecución de grandes proyectos de investigación en Física experimental es un hecho relativamente reciente en México. Estos acontecimientos tuvieron que esperar la institucionalización de las ciencias físicas en nuestro país, misma que se dio ya muy entrado el siglo XX. Cabe recordar que no fue sino hasta el año de 1935 que se funda la ahora extinta Escuela Nacional de Ciencias Físico-Matemáticas donde se ofrecieron por vez primera las licenciaturas de Física y de Matemáticas. Poco tiempo después, en 1938, se fundaría el Instituto de Física de la UNAM, institución propiamente dedicada a la investigación en esta rama del saber y que sería una de las precursoras de proyectos de gran alcance en México[1]. Es alrededor de estos años fundacionales que varios acontecimientos internacionales -ocurridos antes y durante la Segunda Guerra Mundial y en los años siguientes de la "Guerra Fría"- gestarían el desarrollo de programas importantes. Uno de ellos fue el de la "Ultracentrífuga Transelástica", proyecto que fuera desarrollado durante varios años en el Instituto Nacional de Energía Nuclear de México (INEN) y su sucesor el Instituto Nacional de Investigaciones Nucleares (ININ). En el presente artículo explicaremos las razones científicas que motivaron la gestación del proyecto mencionado y narraremos su desarrollo, mismo que culminaría con una extraordinaria ultracentrífuga cuyos diseñadores y constructores fueron distinguidos con el Premio Nacional de Ciencias 1973.[2]

Para nuestros propósitos el presente trabajo se encuentra dividido en diez partes incluyendo esta sección donde ofrecemos la sinopsis del manuscrito. En la segunda sección nos ocuparemos de describir los antecedentes que dieron surgimiento al proyecto de la ultracentrífuga. Por tal motivo debemos remontarnos a la época entre las dos guerras mundiales. Es durante este período que los Estados Unidos idearon la "*Política de Buena Vecindad*". Esta política tuvo -como señalaremos- impacto en la formación de la naciente comunidad institucionalizada de físicos mexicanos. En la tercera sección veremos cómo, a raíz de esta nueva actitud norteamericana, un grupo de científicos radicados en Massachusetts aprovechó las oportunidades que esta nueva medida ofrecía, para brindar apoyo a sus colegas latinoamericanos. Este apoyo se tradujo en el ofrecimiento de becas para estudiantes y especialistas así como en visitas de científicos norteamericanos a la región. En la cuarta sección observaremos que entre los académicos distinguidos que visitaron Latinoamérica se encontraría el matemático George David Birkhoff quien aprovechó la ocasión para exponer una teoría alternativa a la Relatividad General de Einstein. La nueva teoría despertó el interés de varios reconocidos físicos mexicanos quienes, como veremos en la quinta parte de este artículo, la desarrollaron ulteriormente. En la sexta sección examinaremos las consecuencias de una predicción teórica basada en la teoría de Birkhoff, hecha a principios de los sesentas por James Clark Keith, ("Jimmy") joven investigador de la Universidad de Detroit, Michigan. Keith predijo un supuesto "efecto observable" en cuerpos materiales girando muy rápidamente. El éxito en la observación de este efecto confirmaría, según Keith, la veracidad de la teoría de Birkhoff sobre la de Einstein. Para confirmar el efecto, Keith y colegas construyeron en Detroit pequeñas versiones de centrifugas con el apoyo financiero de las fuerzas armadas norteamericanas. Posteriormente al ser retirado el financiamiento militar, Keith se trasladó a México donde hizo estudios doctorales en la UNAM y a la postre laboró en el Centro Nuclear de la Comisión Nacional de Energía Nuclear CNEN (después transformada en INEN y después ININ) donde en colaboración con un grupo de investigadores mexicanos continúo con el proyecto iniciado en Detroit. Ya estando laborando en México en la CNEN, Keith y sus colaboradores mexicanos desarrollaron una ultracentrífuga capaz de levitar magnéticamente una esfera de acero y hacerla girar muy rápidamente al grado que la fuerza centrífuga que actuaba sobre la esfera al girar, llegaba a deformar plásticamente la misma, al rebasar dicha fuerza el límite elástico del acero.

Volviendo al contenido del presente trabajo, en la séptima sección presentamos algunas particularidades importantes sobre la separación isotópica. En la octava describiremos algunos pormenores del desarrollo de la ultracentrífuga en México, en la novena señalaremos cual fue el desenlace del proyecto y en la última sección expondremos nuestros comentarios finales.

2. **La política de la buena vecindad**.

El sábado 4 de marzo de 1933, en la capital de los Estados Unidos, Franklin Delano Roosevelt pronunció ante su nación el discurso de toma de posesión a la presidencia. En él decía:

> *"En el terreno de la política mundial, estableceré en esta nación la política del buen vecino –aquel que tenazmente se respeta a sí mismo, y al hacerlo respeta los derechos de los otros- un vecino que respeta sus obligaciones y respeta el cumplimiento de sus acuerdos dentro y con un mundo de vecinos"*[3]

Con estas palabras Franklin Roosevelt (presidente de 1933 a 1945) apartaba su discurso político de la interpretación imperialista de la Doctrina Monroe hecha por su predecesor Theodore Roosevelt (presidente de 1901 a 1909). En efecto, en 1904 el presidente Theodore Roosevelt había declarado, en su alocución anual sobre el estado de la nación, el derecho de intervención de los Estados Unidos en cualquier nación latinoamericana que flagrante y sistemáticamente cometiera actos que "*a juicio de Estados Unidos*" fueran delictivos.[4] Tres décadas más tarde los Estados Unidos darían un "golpe de timón" en su política hacia Latinoamérica ante su inminente enfrentamiento contra las potencias del Eje. Este cambio de actitud se tradujo en el aparente abandono de la política imperialista del "*big stick*" por la política del "*good neighbor*". Esta supuesta transformación estaba motivada en parte por el temor a tener que cuidarse las espaldas de sus vecinos latinoamericanos en caso de guerra. La estrategia norteamericana vendría acompañada por acciones tendientes a minimizar las tensiones entre Washington y los gobiernos de la región[5]. Una de las acciones seguidas por el sector oficial norteamericano, siguiendo la política de "*buen vecino*", fue la de crear en agosto de 1940 un organismo dependiente de la Secretaría de Estado nombrándolo la *Office for Coordination of Economic and Cultural Relations between the American Republics,* que más tarde tomó el nombre de *Office of the Coordinator of Inter-American Affairs*[6] cuyo objetivo era apoyar al esfuerzo bélico norteamericano, relacionando los intereses de las repúblicas de América Latina a aquellos de los Estados Unidos. En aquel tiempo, fue nombrado coordinador de dicha oficina Nelson Rockefeller quien a su vez invitó a Henry Allen Moe a dirigir la sección a cargo de los intercambios culturales con las repúblicas latinoamericanas. Henry Allen Moe en aquel entonces fungía como Secretario de la Fundación Guggenheim, de Nueva York. Dicha fundación apoyaba económicamente desde 1925, a intelectuales y científicos canadienses y norteamericanos. Hacía 1931 la oficina extendió su programa de becas a latinoamericanos[7]. Este viraje en la política exterior estadounidense no tardó en ser aprovechado por un grupo de científicos norteamericanos que simpatizaban con las repúblicas latinoamericanas.

### 3. Las tertulias de Cambridge, Massachusetts.

Hacia fines de 1941 en la ciudad de Cambridge, Massachusetts, se reunían informalmente un grupo de académicos de esa localidad, animados por la nueva política de "*buena vecindad*". Entre ellos se encontraban, los matemáticos Norbert Wiener de MIT, George D. Birkohff de Harvard, el astrónomo Harlow Shapley del Observatorio de Harvard, el fisiólogo Walter B. Cannon de Harvard y el físico mexicano Manuel Sandoval Vallarta de MIT. El propósito de sus tertulias era el discutir cómo lograr el establecimiento más firme de relaciones científicas con sus colegas latinoamericanos, aprovechando el cobijo de la nueva política de la Casa Blanca. En estas reuniones salió a relucir que todos ellos tenían algún tipo de contacto con el ya mencionado Henry Moe. Los vínculos entre Moe y cada uno de los miembros de la tertulia se habían establecido por diversas razones: ya sea porque algunos de ellos habían actuado como referentes en comités de becas de la propia fundación Guggenheim de la cual, como se mencionó, Moe era Secretario, o bien poseían lazos personales de amistad con él y además tenían contactos en los países Latinoamericanos. Por otro lado, uno de ellos (Harlow Shapley) había establecido, como veremos a continuación, una buena amistad con Luis Enrique Erro, quien fuera un brillante impulsor de la astronomía mexicana.

Viene al caso relatar cómo se gestó la relación amistosa entre Erro y Shapley. Luis Enrique Erro era miembro del Consejo Nacional de la Educación Superior y la Investigación Científica (CONESIC) que fuera fundada en 1935 durante la administración del presidente Lázaro Cárdenas. Uno de los objetivos del CONESIC era el dar impulso a la investigación científica en el país. Desde este espacio Erro concibió la creación de un moderno observatorio astrofísico para México[8]. A finales de la década de los treintas, Erro formó parte del cuerpo diplomático mexicano: primero en París, Francia, y luego en Boston, Massachusetts. En Boston dedicó parte de su tiempo libre a la que era su afición, las observaciones astronómicas, esto a pesar de no tener estudios formales en astronomía. Allí se reunió con miembros de la Asociación Americana de Observadores de Estrellas Variables (AAVSO, por sus siglas en inglés) y trabó amistad con León Campbell quién era el coordinador de dicha asociación, y que además trabajaba en el observatorio astronómico de la Universidad de Harvard. Campbell se encargaría de poner en contacto a Luis Enrique Erro con Harlow Shapley director del observatorio y con otros astrónomos de Harvard. Hecho el contacto entre ambos, el don de gentes de Erro y el interés de Shapley por Latinoamérica pronto capturó la amistad entre este par de personajes. Ya en confianza Erro le refirió a Shapley su plan de proponer al gobierno mexicano la construcción de un moderno observatorio astrofísico, además de hacerle saber de su amistad con Lázaro Cárdenas y con el que sería el próximo presidente del país, Manuel Ávila Camacho. Por otro lado, la idea de tener acceso a un observatorio a latitudes al sur de la frontera de Estados Unidos prendió el interés de Shapley y de otros astrónomos de Harvard. De regreso en México, Erro concretó los apoyos necesarios para la realización del proyecto de construir un observatorio, contando además con la valiosa ayuda que recibiría del observatorio de Harvard, principalmente, en el diseño, construcción e instalación de un telescopio tipo Schmidt[9]. El financiamiento del observatorio sería proporcionado por el gobierno mexicano y por las fundaciones Rockefeller y Jenkins[10]. Erro actuaría como el primer director del nuevo observatorio. Como hecho importante debemos mencionar que durante su estancia en Harvard, Erro conoció a Carlos Graef Fernández, joven físico becario Gugenheim que estudiaba su doctorado en el MIT, bajo la dirección de Manuel Sandoval Vallarta y además había recibido cursos de astrofísica en Harvard. De regreso a México Erro invitaría a Carlos Graef a colaborar en el futuro observatorio, en calidad de director asistente.

Finalmente la inauguración oficial del Observatorio de Astrofísica de Tonantzintla se programó el 17 de febrero de 1942. A la ceremonia asistieron diversas personalidades del gobierno mexicano, entre ellas el presidente de la república, Ávila Camacho. Además se acordó celebrar después de la ceremonia, un congreso internacional de astrofísicos en el mismo observatorio de Tonantzintla y posteriormente en la cercana ciudad de Puebla.[11] En el fondo, la intención del congreso poblano era resaltar el éxito de la política de "*buena vecindad*" reuniendo lo más selecto de los astrofísicos norteamericanos con la naciente comunidad de científicos mexicanos. Es de notar que unas semanas antes del ataque a Pearl Harbor (7 de diciembre de 1941), Harlow Shapley recibió una nota escrita por Henry Wallace, vicepresidente de los Estados Unidos donde se le hacía saber que al Presidente Roosevelt le complacería que los científicos estadounidenses aceptaran la invitación a la ceremonia de inauguración, "*hubiera o no guerra*"[12]. Esto demuestra el gran interés que el gobierno norteamericano tenía por la participación al evento de los científicos norteamericanos invitados. Uno de esos invitados sería el distinguido matemático George David Birkhoff que como ya hemos mencionado asistía a las tertulias de Cambridge.

### 4   George David Birkhoff

George David Birkhoff (1884-1944) gozaba en los años 40 de la fama de ser el más famoso matemático norteamericano de su tiempo. Su fama derivaba entre otras cosas de su trabajo en la teoría ergódica[13]. Mediante la aplicación de la medida de Lebesgue, probó lo que ahora se conoce como "teorema ergódico" (1931-32) mismo que permitió formalizar la teoría cinética de los gases de Maxwell-Boltzmann. Su popularidad, también se debía al hecho de haber resuelto el último teorema geométrico de Poincaré (en 1913), además de su notable trabajo en sistemas de ecuaciones diferenciales.

Hacia 1940 tendría lugar en Harvard el Congreso Internacional de Matemáticas cuyo organizador era Birkhoff, sin embargo esta reunión debió suspenderse debido a la imposibilidad de viajar de los matemáticos europeos y asiáticos a los Estados Unidos ante el inminente estallido de la Segunda Guerra Mundial. Entre los congresistas que asistirían a ese malogrado congreso se encontraba invitado el matemático peruano Godofredo García quien mantenía, hacía tiempo, intercambio epistolar con Birkhoff. Por su parte, García había propuesto, en julio de 1939 a la Academia de Ciencias Exactas, Físicas y Naturales de Lima (de la que era fundador y Presidente), la designación de Birkhoff como académico asociado.[14]

La cancelación del congreso matemático en Harvard organizado por Birkhoff le hizo considerar un cambio de planes e ir al Perú a recibir su nombramiento como académico. Por su parte Harlow Shapley y Sandoval Vallarta al enterarse del plan de Birkhoff de visitar el Perú, gestionaron una invitación para este último al congreso de Tonantzintla. Birkhoff consideró esta proposición como una oportunidad para presentar su teoría alternativa a la gravitación de Einstein ante un público compuesto por lo más selecto de los astrofísicos norteamericanos. La asistencia de éstos estaba asegurada por el apoyo presidencial de Roosevelt "*hubiera o no guerra*".

Tras las invitaciones recibidas, Birkhoff comunicó a Moe su interés por realizar una gira en Latinoamérica invocando la política norteamericana de "*buena vecindad*" sugiriendo que la oficina de Nelson Rockefeller podría estar interesada en prestarle apoyo y contribuir al financiamiento de su gira. Moe lo apoyó sin reservas. De esta manera Birkhoff obtuvo respaldo financiero de la oficina de Rockefeller y de la Fundación Guggenheim. La gira planeada cubriría ciudades de México, Chile, Argentina, Uruguay y Perú.

## 5 La Teoría de Birkhoff y su desarrollo en México.

En febrero de 1942 Birkhoff presentó una síntesis de su teoría en el marco del congreso de astrofísica[15]. Su propuesta contemplaba a la Gravedad como una fuerza de carácter tensorial actuando en el espacio plano de Minkowski. El modelo daba como válida la teoría de la Relatividad Especial de Einstein. Lo interesante de este modelo es que en él se consideraba a la Gravedad como una fuerza derivada de un potencial gravitacional, de la misma manera como se consideran otras fuerzas de la naturaleza, como por ejemplo la fuerza electromagnética derivada de un potencial electromagnético. Para ambos potenciales –en la teoría de Birkhoff- el espacio en el que actúan es simplemente un espacio plano "pasivo".

Esta idea está en clara contraposición conceptual con la teoría de la Relatividad General de Einstein en donde la interacción gravitacional no es propiamente una fuerza sino es una consecuencia del efecto causado por la geometría del espacio. A pesar de dicha contraposición, la simplicidad conceptual de la Teoría de Birkhoff y la posibilidad de tratar a todas las fuerzas de la Física dentro de un mismo espacio plano, atrajo la atención de varios de los asistentes a las pláticas de inauguración de Tonantzintla. Carlos Graef recuerda con estas palabras su impresión de aquella reunión,

> *"Dos científicos mexicanos se entusiasman con las ideas de Birkhoff e inician investigaciones para obtener las consecuencias de su teoría. Se trata de los doctores Alberto Barajas y Carlos Graef Fernández. Otro científico estadounidense que acudió a Puebla al Congreso fue el doctor Jesse W. Beams, <u>el inventor de las ultracentrífugas</u> de suspensión magnética"*[16]. (El subrayado es nuestro, más tarde hablaremos de Beams)

Posteriormente en 1943 Birkhoff extendería su teoría para incluir ya no solo el potencial gravitacional, sino además incorporaría en ella el potencial atómico y el electromagnético[17]. En este trabajo Birkhoff plantea que las señales gravitacionales se transmiten a la velocidad de la luz en un hipotético fluido perfecto, incompresible y de viscosidad nula que llena el espacio plano de Minkowski. Este trabajo precisa lo siguiente,

> *"El fluido perfecto se distingue por el requerimiento…de que la velocidad de perturbación tiene que ser la de la luz para todas las densidades, con la correspondiente ecuación de estado $p = \rho/2$, donde $\rho$ y $p = f(\rho)$ designan respectivamente la densidad y la presión del fluido".*[18]

Y más adelante afirma,

> "*Suponemos que la materia es: o bien ese fluido especial, homogéneo adiabático, matemáticamente satisfactorio para la cual $p = \rho/2$ o bien, presumiblemente cualquier forma de materia en la cual <u>la velocidad de perturbación es aquella de la luz bajo todas las circunstancias</u>*"[19] (el subrayado es nuestro).

Estas afirmaciones levantaron inmediatamente críticas en el ámbito internacional pues el modelo de "bolas duras" de Birkhoff ($p = \rho/2$) suponía explícitamente que la velocidad de una perturbación -como podría ser la del sonido- coincide con la velocidad de la luz[20]. Aunque es muy pertinente aclarar que la teoría nunca fue desarrollada en presencia de campos de materia y por lo tanto no se puede afirmar a ciencia cierta si esta aseveración es correcta. Por otro lado la equivalencia entre la masa inercial y gravitacional ya no surgía de una manera "natural" como en la teoría de Einstein, esto es, como consecuencia de la geometría del espacio sino como un regla de la teoría de Birkhoff. Naturalmente no tardaron en surgir objeciones al respecto. Fue en ese mismo año de 1943 que Hermann Weyl califica la nueva teoría como una representación "*degenerada*" de la teoría de Einstein aplicada a campos gravitacionales débiles. En sus comentarios Weyl precisa que es sencillo romper la relación entre inercia y gravitación en la teoría de Einstein simplemente despreciando términos de orden superior en algunos desarrollos en serie para el caso de campos gravitacionales débiles[21]. Por lo tanto Weyl aseguraba que la nueva teoría de Birkhoff resultaba ser un caso limitado de la de Einstein y que además ratificaba arbitrariamente el principio de equivalencia de masas.

El primero en contestar las críticas de Weyl fue Alberto Barajas quien respondió señalando que la teoría de Birkhoff no es en realidad una forma degenerada de la de Einstein para campos débiles. Para mostrarlo hace una detallada comparación entre ambas teorías alegando que verdaderamente la de Birkhoff es una teoría alternativa que parte de conceptos distintos.[22] La respuesta de Weyl a Barajas no se hace esperar. Podemos resumir sus críticas en cuatro principales puntos: 1) En la teoría de Birkhoff, se disuelve la conexión entre la métrica y la gravitación 2) La proporción entre la masa inercial y la gravitacional se convierte en un misterio como en los tiempos anteriores a Einstein 3) El fluido perfecto de Birkhoff aparece como una primitiva entidad física irreducible 4) No parece haber indicación de que las ecuaciones de la mecánica surjan de una ley universal de conservación de la energía y momento. Aunado a las cuatro objeciones que acabamos de enumerar debemos señalar que el artículo de Weyl está escrito en un tono provocador llegando a calificar abiertamente la teoría de Birkhoff como retardataria pues dice,[23]

> *"Cualquier teoría que rompa la unidad de inercia y gravitación, como lo hace la reciente teoría de Birkhoff de 'gravitación en un mundo plano', <u>nos retrocede a una postura anterior a Einstein donde tendríamos que aceptar sin entender, la identidad de la masa y el peso</u>"*.[24] (El subrayado es nuestro)

En esa misma época Birkhoff se hallaba en México trabajando con Barajas, Graef y Sandoval Vallarta. Entendiblemente ellos unen sus esfuerzos para dar respuesta a todas y cada una de las críticas hechas por Weyl publicando en coautoría un artículo[25].

En dicha publicación la primera objeción hecha por Weyl es recusada argumentando que la Relatividad General de Einstein abandona "*la descripción tradicional de la naturaleza*" en términos de cuatro variables fundamentales independientes (coordenadas espaciales y la temporal). Arguyen que lo que se hace en la de Birkhoff es volver a adoptar esa descripción tradicional, lo que a la postre permitirá unificar más fácilmente las fuerzas de la naturaleza. Sobre la segunda objeción los autores argumentan que Weyl malinterpreta la teoría de Birkhoff y dicen que si existe una equivalencia entre masas (inercial y gravitacional) la cual surge del hecho de que en las ecuaciones gravitacionales, el tensor de energía tiene una dependencia lineal con la densidad de masa. La tercera objeción es saldada diciendo que efectivamente, el "*fluido perfecto*" es el único postulado físico de la teoría de Birkhoff. La última objeción es considerada al igual que la segunda, como una malinterpretación de Weyl, pues afirman que tanto la conservación de energía como la de momento son consecuencia de las ecuaciones de movimiento de la teoría de Birkhoff y viceversa. En una publicación posterior, Graef se encargaría de atender detalladamente este punto[26].

Birkhoff ampliaría su modelo en agosto de 1943 tomando en cuenta además las interacciones electromagnéticas.[27] Unos meses después Birkhoff fallece el 12 de noviembre de 1944 y a partir de entonces sus principales seguidores, Carlos Graef y Alberto Barajas extendieron sistemáticamente su teoría y animaron a algunos de sus colegas como Marcos Moshinsky[28] y Fernando Alba Andrade[29] a incursionar en el tema explorando las posibilidades de esta nueva teoría[30]. Por otro lado Sandoval Vallarta abandonaría la cruzada emprendida por Graef y Barajas pues escribiría tan solo un artículo más al respecto[31]. En este interesante artículo Sandoval Vallarta presenta una tabla comparativa entre la relatividad general y la teoría de Birkhoff. En dicha tabla es fácil analizar los puntos de concordancia y discrepancia entre ambos modelos. En la comparación señala que ambas teorías son matemáticamente congruentes y con ambas se puede pronosticar, ajustando ciertos parámetros las tres pruebas clásicas de la relatividad. No obstante Sandoval Vallarta aclara que "…*es el experimento el que deberá decidir sobre sí la teoría de Birkhoff es aceptable o no*"[32]. Sandoval Vallarta no volvería a escribir más sobre el tema. En contraste Carlos Graef publicaría entre 1944 y 1968 un total de 19 artículos sobre la teoría de Birkhoff abarcando diversos aspectos como: la dinámica entre dos cuerpos bajo atracción gravitacional, el principio de conservación de la energía, el campo gravitacional producido por una masa puntual en movimiento arbitrario y muchos otros temas relativos al modelo de Birkhoff. [33]

Cabe aquí abrir un paréntesis para mencionar que la actividad de Cralos Graef no se limitó únicamente a la investigación científica sino que además de ser un gran maestro fue un entusiasta funcionario promotor de la energía nuclear en México junto con Nabor Carrillo Flores, Alberto Barajas y Sandoval Vallarta.[34] Para las intenciones de este trabajo como veremos más adelante tan solo es pertinente señalar algunas de las actividades de Graef en relación a sus actividades de promoción

En 1960 Graef había sido electo "Gobernador Representante de México ante la junta de Gobernadores del Organismo Internacional de Energía Atómica (OIEA)" y gracias a su activa gestión se firmó un convenio el 18 de diciembre de 1963 entre la OIEA y México cuyo propósito era "dotar a México de un reactor de investigación nuclear, así como de efectuar una cesión de uranio para tal reactor" [35] Este hecho es relevante por dos motivos: el primero es que con la adquisición de un reactor, México entraría de lleno a la era de los reactores nucleares pero el segundo hecho es que dicha "cesión de uranio" (enriquecido) implicaba establecer una dependencia del país con el extranjero, ya que México no contaba la tecnología para el enriquecimiento del uranio. De lo anterior Graef estaba consciente y realizó esfuerzos para eliminar la dependencia ya que más tarde, durante su gestión como Director del Centro Nuclear de México, actuó en consecuencia. En efecto, en la memoria de labores 1968-1969 del Centro podemos leer: *"… ante la posibilidad próxima de que México seleccione un primer reactor de potencia , dentro de sus actuales posibilidades, se han impulsado las investigaciones […] para que puedan fabricarse en nuestro territorio combustibles nucleares"*.[36] Como veremos más adelante a comienzos de los setentas se exploraría en el Centro Nuclear el método de enriquecimiento del uranio por ultracentrifugación. Esto se generó debidos a las circunstancias que comenzaron a gestarse dos décadas antes y que relataremos a continuación.

### 6   Jimmy Keith y las centrífugas.

México no fue el único lugar donde la teoría de Birkhoff despertó interés. A finales de los años 50 un pequeño grupo independiente en la Universidad de Detroit, Michigan laboraba en un programa conocido como ERG (Experimental Research into Gravity) cuya finalidad era verificar experimentalmente la teoría de Birkhoff[37]. En dicho programa trabajaban: Donald J. Kenney como responsable, John D. Nixon al cuidado de la ingeniería electrónica y James Clark Keith a cargo del desarrollo de los aspectos teóricos. El programa ERG de la Universidad de Detroit, según testimonio del propio Keith, fue inspirado por Carlos Graef y Jesse W. Beams[38]. Cabe recordar que Beams (el padre de las ultracentrífugas de suspensión magnética) y Graef, como ya lo hemos mencionado, asistieron a la presentación inaugural que hizo Birkhoff de su teoría en Tonantzintla. Es posible que a raíz de este encuentro entre ambos se hayan gestado algunas ideas sobre el cómo probar la teoría de Birkhoff mediante el uso de centrifugas, por lo que no debe extrañarnos que Keith les atribuya la "inspiración" del programa ERG.

Sobre James Clark Keith debemos señalar que antes de integrarse al grupo de la Universidad Detroit, "Jimmy" -como se le conocía- había hecho sus estudios de licenciatura en Física en el MIT donde tomó algunos cursos impartidos por Garret Birkhoff (hijo de George David) de la Universidad de Harvard y por Stanley Deser de la vecina Universidad de Brandeis. Bajo la tutela de estos profesores y de su director de tesis, el distinguido matemático de origen holandés Dirk Jan Struik[39], Keith se familiarizó con la teoría de la gravitación de Birkhoff (padre) de manera que en 1957 se tituló de licenciatura con la tesis "Sobre la teoría de Birkhoff de la materia y fuerza en el espacio-tiempo de Minkowski" [40]. Hacia el año de 1958 Jimmy se encontraba ya trabajando en el programa ERG de la Universidad de Detroit.

Para verificar la teoría de Birkhoff, Jimmy hizo sus propios cálculos teóricos- sustentados en la mencionada teoría- para con ellos proponer una prueba experimental que supuestamente la confirmaría. Para tal fin el grupo ERG construyó una pequeña centrífuga cuyo rotor era una esferita de acero la cual se hacía girar a gran velocidad, para después retirarle la fuerza que había impulsado su giro y posteriormente medir cuanta energía cinética iba perdiendo la esferita al dejarla rotar libremente a la deriva. Esta pérdida que supuestamente se observaría, sería debida a la interacción entre la materia en rotación y la reacción de fuerza retardada producida por el campo gravitacional de Birkhoff. Cabe precisar que este efecto no está presente en la teoría General de Relatividad de Einsstein.

La esferita se mantenía suspendida magnéticamente y en una pequeña cámara al vacío, esto con el propósito de minimizar efectos de fricción con el aire, que tenderían a frenar al rotor. El trabajo específico de Jimmy fue calcular cual sería el decremento de la energía cinética de un rotor al girar a la deriva en el marco de la teoría de Birkhoff[41]. Para tal efecto Jimmy calculó primero la pérdida de potencia $P$ de un rotor que interacciona con el Universo,

$$P = k\, k'E_{kin}\left(\frac{a\omega}{c}\right)^3 \omega\ ,$$

donde $\omega$ es la velocidad angular del rotor, $k$ es una constante (para una barra homogénea en rotación $k = 2$ y para un rotor esférico $k = (75\pi/128)$, $a$ es el máximo radio de giro, $E_{kin}$ es la energía cinética del rotor, $c$ es la velocidad de la luz, $k'$ es una constante que representa "*la reacción del sistema con el Universo*". La constante $k'$ en la teoría de Birkhoff es igual a la unidad. Posteriormente haciendo uso de la ecuación anterior, Jimmy calculó cuál debía ser la razón del decremento $\Delta = df/dt/f$ de la frecuencia de giro $f$ para un rotor esférico girando libremente. El resultado que obtuvo fue,

$$\Delta = \frac{df}{dt}/f = -\frac{1}{2}k(a\omega/c)^3\omega \qquad \text{con } \omega = 2\pi f \qquad (1)$$

A partir de esta relación Jimmy hizo una valoración de la magnitud experimental del efecto, llegando a la conclusión de que el efecto es muy pequeño, pues aun para un hipotético rotor esférico de acero girando a frecuencias del orden de megahercios el valor esperado del decremento de la frecuencia $\Delta$ corresponde a 3 ciclos por segundo por día. Esto implica tener un sistema de detección de frecuencia capaz de distinguir 3 partes en $10^6$. Además señala que este decremento hay que distinguirlo experimentalmente de otros efectos presentes que frenan el libre giro del rotor, tales como las torcas magnéticas y el efecto Coriolis[42]. En pocas palabras el efecto a medir según los cálculos de Jimmy era muy pequeño y podría estar enmascarado por otros. No obstante, con este reto enfrente el grupo ERG de Detroit estimó que era posible medir este pequeño decremento en la frecuencia de giro del rotor y para tal fin como ya hemos señalado, construyó una centrifuga del tipo desarrollado por Jesse W. Beams

Para obtener el financiamiento requerido para su proyecto, el grupo ERG se aprovechó de una situación muy peculiar que se dio en el seno de la sociedad norteamericana derivada del período llamado de la "*guerra fría*". El 4 de octubre de 1957 la URSS puso en órbita el *Sputnik 1*, el primer satélite artificial. El público norteamericano tomó esta noticia con mucha aprensión y su gobierno actuó en consecuencia incrementando su inversión en investigación que pudiese mantener a los Estados Unidos en el liderazgo tecnológico[43]. En efecto, en 1958 el Departamento de Defensa de los Estados Unidos creo la agencia ARPA (Advanced Research Projects Agency) cuyo objetivo era *"prevenir la sorpresa tecnológica, como la del Sputnik…pero también crear sorpresa tecnológica para nuestros enemigos"* mediante el apoyo a proyectos *"innovadores de tecnologías radicales"*[44]. Aprovechando esta situación el grupo ERG solicitó a la Army Research Office con sede en Durham, Carolina del Norte, (AROD) apoyo financiero para su proyecto. El apoyo a dicho proyecto fue aprobado y se acordó que sería supervisado por USA TRECOM (United States Transportation Research Command). Resulta interesante preguntarnos porqué TRECOM, institución encargada de coordinar la investigación en nuevos medios de transporte de personal y equipo militar, supervisó dicho proyecto. Tal vez el ejército americano debió pensar que los conocimientos derivados del proyecto podrían de alguna manera ser utilizados para manipular los efectos de la gravedad y así facilitar la transportación de vehículos militares.[45] Sin embargo ante la proliferación de financiamiento a proyectos cuyo objetivo era hipotético, el Congreso de los Estados Unidos comenzó a mediados de los sesenta a cuestionar el financiamiento militar a proyectos que tuviesen una dudosa relación con objetivos castrenses y en 1969 aprobó una ley (The Mansfield Amendment, Public Law 91-121) prohibiendo completamente dicha subvención[46]. Para el caso particular del grupo ERG, como era de esperarse, el peculio fue interrumpido y a la postre el grupo cesó sus actividades en 1964. Fue entonces (1968) cuando Jimmy se trasladó a México por invitación de Carlos Graef para realizar bajo su supervisión, el doctorado en Física en la UNAM[47]. Después de doctorarse en 1970, Jimmy fue contratado en 1971 como "investigador especial" del Centro Nuclear de México. En aquel entonces Graef era el Director General de esa institución donde recién había creado el departamento de ultracentrífugas. Allí, Jimmy encontraría el apoyo y los recursos suficientes para construir una ultracentrífuga y así podría reanudar su búsqueda del efecto gravitacional por él predicho y por otro lado- como veremos- podría explorar la posibilidad de encontrar una nueva técnica de separación de isótopos de uranio. En efecto, la misión "oficial" del departamento era hacer *"la investigación fundamental y aplicada para determinar la factibilidad física e industrial de difusión ultrarápida intermetálica, enfocada sobre la separación de isótopos de uranio por centrifugación en estado sólido"*[48]. En simples palabras el objetivo del departamento recién creado era investigar el método de enriquecimiento de uranio por centrifugación, pero "en estado sólido". Para tal fin se centrifugarían esferitas de acero con alma de uranio metálico. Conjuntamente el proyecto implicaba tener la capacidad de monitoreo continuo de la velocidad de rotación de dichas esferas al ser centrifugadas. Esto último tal vez permitiría observar y medir el efecto vaticinado por los cálculos de Jimmy. En pocas palabras el proyecto prometía el logro simultáneo de dos objetivos por un lado la continuación de la búsqueda del efecto predicho por Jimmy y por otro lado, algo muy importante, la posibilidad de enriquecer uranio, lo que daría independencia a la naciente industria nuclear mexicana.

## 7  El Enriquecimiento de Uranio por Centrifugación.

El uranio en forma natural se presenta como una mezcla de tres isótopos: $^{234}$U (234.11.41 uma) 0.0058%, $^{235}$U (235.1175uma) 0.711% y $^{238}$U (238.1252 uma) 99.283 % en peso. La abundancia del primero es tan pequeña que éste puede despreciarse para fines prácticos. La finalidad del proceso de trasformación de uranio natural en enriquecido es incrementar la concentración del isótopo ligero $^{235}$U. Esta pequeña concentración del 235 en el uranio natural puede incrementarse a través de un proceso de separación de isótopos. La importancia de la separación del isótopo 235 reside en que su fisión constituye el mecanismo básico para la producción de energía nuclear además de ser el único isótopo en cantidad apreciable existente en la naturaleza que es fisionable mediante neutrones térmicos. De allí radica la importancia estratégica de dominar el método de enriquecimiento del uranio y un país como México, en aquel tiempo con un insipiente programa nuclear, debía incursionar en esta importante área, por lo que Carlos Graef no dudo en apoyar el proyecto de enriquecimiento por centrifugación.

El uso de centrífugas para la separación isotópica había sido propuesto por vez primera por los británicos Frederick A. Lindemann y Francis W. Aston poco después del descubrimiento de la existencia de isótopos de elementos ligeros, hecho en 1919 por el propio Aston en el Laboratorio Cavendish en Inglaterra[49]. Durante las siguientes décadas, varias centrífugas primitivas en varios laboratorios del mundo fueron empleadas sin éxito para tratar de lograr separaciones isotópicas en medios gaseosos. Sin embargo, no fue sino hasta 1936 que Jesse W. Beams y su grupo de la Universidad de Virginia lograron separar isotopos de cloro[50]. A pesar de su éxito, el diseño de Beams tenía el serio inconveniente de que su aparato utilizaba muchos kilowatts eléctricos debido a la gran fricción mecánica que se presentaba en los cojinetes que sostenían al eje del rotor de la centrífuga.

El estallido de la Segunda Guerra Mundial concentró al grupo de Beams dentro del proyecto Manhattan en el problema de la separación isotópica del Uranio por medio de centrifugación gaseosa. Otro método de separación alternativo contemplaba la difusión gaseosa a través de tamizas. La responsabilidad dentro del proyecto Manhattan, de explorar la factibilidad del mencionado par de métodos de separación recayó sobre Harold Urey (Nobel 1934) quien optó por el método de difusión gaseosa en lugar de centrifugación gaseosa, para obtener el suficiente Uranio enriquecido en el desarrollo de la bomba atómica[51]. Por lo tanto, el proyecto de enriquecimiento por centrifugación fue interrumpido en 1943.

Diez años más tarde en 1953, la Comisión de Energía Atómica de los Estados Unidos decidió revivir el programa de separación por centrifugación al advertir que un grupo de la Universidad de Bonn en Alemania dirigido por Wilhelm Groth afirmaba que la separación por centrifugación podía ser más económica que la separación por difusión gaseosa. Cabe señalar que Wilhelm Groth había presidido durante la segunda guerra mundial el proyecto de enriquecimiento de uranio por centrifugación gaseosa para la Alemania nazi. Además en ese mismo año de 1953 el Almirante brasileño Álvaro Alberto se había reunido con Groth, Otto

Hahn y Paul Harteck en el Instituto de Física de Hamburgo para arreglar el envío al Brasil de 3 centrífugas para la separación isotópica de uranio[52].

Temiendo perder el monopolio de enriquecimiento de uranio, la Comisión de Energía Norteamericana revivió el programa de centrífugas en septiembre de 1954. Para entonces F.T. Holmes, miembro del grupo de Beams, ya había subsanado el problema del alto consumo de energía por perdidas por fricción de las centrifugas al levitar magnéticamente los rotores de las mismas, eliminando así los cojinetes mecánicos[53]. Con el dominio de la levitación magnética de los rotores, esto es sin el empleo de un eje mecánico de giro, el grupo de Beams consiguió confinar los rotores en cámaras de muy alto vacío eliminando así -en gran medida- la fricción producida por el aire circundante sobre el rotor. De esta manera Beams y Holmes lograron hacer girar pequeñas barras de acero a grandes velocidades hasta llegar a las mil revoluciones por segundo[54]. Incidentalmente debemos mencionar que este logro permitió al grupo de Beams emprender nuevas investigaciones. Por ejemplo, uno de sus experimentos consistía en poner a girar un rotor a alta velocidad para después dejarlo libremente girar a la deriva y así observar su desaceleración, esto con el fin de investigar qué efectos tendrían algunos factores en la desaceleración del rotor. Efectos como la fricción del rotor con aire remanente en la cámara de vacío, efectos de vibraciones mecánicas o bien efectos de frenado por la magnetización del rotor.[55] Además de los mencionados factores, Beams conjuntamente con Graef, inspiraron -según la versión del propio Jimmy- a la Universidad de Detroit para formar el grupo EGR con el objetivo de intentar observar un efecto de frenado gravitacional en los rotores, predicho por Jimmy[56].

Para finalizar esta sección queremos indicar que el método de enriquecimiento de uranio por centrifugación gaseosa es actualmente el proceso comercial más utilizado en el mundo, sin embargo recalcamos que la centrifugación en estado sólido hoy en día se encuentra aún en fases experimentales.[57]

## 8 La Ultracentrífuga Mexicana.

El desarrollo de la ultracentrífuga comenzó en mayo de 1971 en los locales del Centro Nuclear de México en Salazar, Edo. Méx., con Jimmy al frente del departamento de ultracentrífugas. El personal de tiempo completo con el que contaba Jimmy era de: tres jóvenes licenciados en física (Roberto Jiménez Ornelas, Jaime Morales Sandoval y Ofelia Canales del Olmo) y un pasante de ingeniería electrónica (Carlos Cabrera Cruz - coautor del presente artículo), así como del apoyo de seis técnicos en electrónica y en mecánica (Roberto Andrade Muñoz, Edgar Quiroz Miranda, Jaime Juárez Jiménez, Manuel Rodríguez Uribe, Humberto Rivera Meza, José D´ Oporto Zavala) y dos asesores Adriano Di Luca en electrónica y sistemas de conteo y el Ing. Leopoldo Nieto Casas en cuestiones de Ingeniería Mecánica. No bien este grupo había iniciado sus discusiones sobre el diseño y las especificaciones que requeriría el equipo a construir, cuando apareció publicado un artículo en *The Journal of Scientific Instruments* en junio de 1971, esto es un mes después de recién iniciado el proyecto[58]. Esta publicación escrita por Johan K Fremerey de la Universidad de Bonn en Alemania influiría en las especificaciones de diseño que debía tener la ultracentrífuga del Centro Nuclear.[59] En dicho artículo Fremerey reporta la construcción de una ultracentrífuga con la que había medido la razón de decremento Δ de la frecuencia de una esferita de acero girando a la deriva. En su experimento Fremerey midió el frenado de la misma por efecto Coriolis así como por torcas magnéticas. Lo interesante de esta publicación es que el frenado del rotor por estos efectos resultaba ser lineal con la frecuencia de giro ω tal y como lo había predicho Jimmy en su artículo de 1963[37]. Sin embargo Fremerey señalaba que para poder diferenciar una razón de frenado gravitacional proporcional a $\omega^4$ (ver ecuación 1) tal y como Jimmy lo predecía, la frecuencia del rotor de su centrífuga debería ser observada con confiabilidad por al menos durante "*un par de días*" para poderla diferenciar de los otros efectos, cosa que su sistema no lo conseguía. En sus propias palabras,

> "*La fiabilidad* [de mi centrífuga] *deberá ser incrementada hasta tener una estabilidad que dure "un par de días" con el objeto de poder observar un efecto gravitacional, el cual debe manifestarse según las consideraciones teóricas de* [Jimmy] *Keith*"[60]

Como consecuencia de la publicación de Fremerey, para los propósitos de Jimmy, las especificaciones del equipo que se había de construir en el centro Nuclear de México tenían que ser mejores que las del aparato de Femerey, si se quería observar el efecto gravitacional. Esto los obligaba a diseñar y construir una centrífuga que permitiera girar un rotor esférico de acero (de diámetro del orden de centímetros) a frecuencias muy altas (del orden de kilohercios) en un medio al alto vacío ($10^{-8}$ torr) y levitando el rotor en un campo magnético durante largos períodos de tiempo (del orden de una semana) sin que el rotor presentase desequilibrios y la circuitería electrónica exhibiera inestabilidades o desvíos. Al mismo tiempo el sistema debería ser capaz de medir "en línea" la frecuencia de giro del rotor y sus variaciones con gran precisión (del orden de milihercios). Conjuntamente el sistema debía contar con un archivo automático de datos en línea así como de registros y control digitales. Entre tanto el instrumento debía disponer de sistemas periféricos de protección, refrigeración y un sistema de alimentación eléctrica para interrupciones de suministro. Hay que recordar que a principios de los setentas -época en que se diseñó y construyó la primera ultracentrífuga mexicana- la tecnología de microcircuitos y circuitos integrados estaba aún en su juventud. No obstante en menos de tres años el grupo del departamento de ultracentrífugas logró la meta de construir y operar dentro de especificaciones el instrumento y listo para intentar "separación isotópica".

Por otro lado, la idea presentada por Jimmy para la separar isótopos de uranio era muy simple y consistía en centrifugar una esferita de acero a la cual se le debía incrustar en su eje de giro una barrita cilíndrica de uranio metálico. La figura 1 muestra un bosquejo de su propuesta.

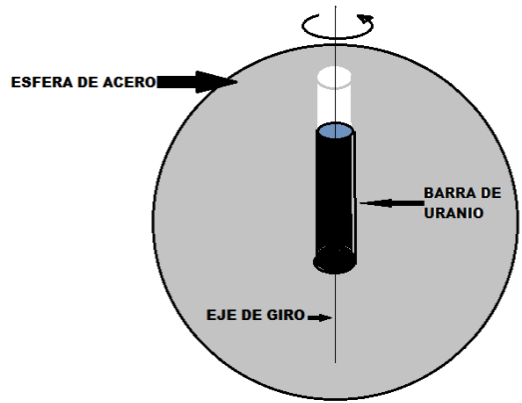

Figura 1. Concepto simple de centrifugación en sólido.

La manera en que Jimmy idealmente concebía la difusión del uranio a través de la esfera de acero es que al centrifugar la esfera, el uranio más pesado se iría más rápido hacia la periferia de la esfera enriqueciendo la parte central con el uranio ligero, al migrar este último más lentamente. Esta idea se ilustra en la figura 2 misma que está tomada de una memoria histórica elaborada en 1982 por el propio departamento de ultracentrífugas. En esa misma memoria podemos leer las suposiciones que hicieron al inicio del proyecto: *El flujo del uranio a través de la esfera de acero resulta del movimiento átomos individuales… Este flujo ocurre a través de un medio fijo… y la difusión forzada domina*"[61]. Como explicaremos más adelante las suposiciones mencionadas no se presentan en la realidad.

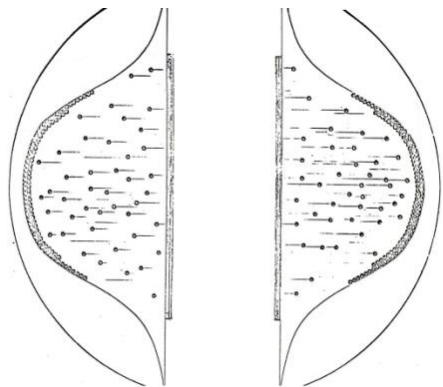

Figura 2. Concepto original de la difusión del uranio.

Sin embargo, la idea del proyecto le debió parecer razonable al Comisión Nacional de Energía Nuclear y el proyecto fue aprobado. Además de que posiblemente tomaron en consideración que ya había antecedentes de separación de isótopos mediante el método del centrifugado pero cabe aquí enfatizar que este método exitoso se empleaba sobre medios gaseosos pero no como en este caso, en estado sólido.

Hacia 1973 la primera versión del aparato se hallaba en operación. El instrumento logrado era único en su clase en el mundo ya que entre otras de sus características, adquiría los datos de frecuencia en tiempo real y los mostraba en forma continua, por lo que la Academia de Ciencias A.C. (de México) otorgó el Premio Nacional de Instrumentación 1973-1974 al grupo de ultracentrífugas[62]. El instrumento original podía girar esferas de acero de hasta una pulgada de diámetro y versiones posteriores eran capaces de girar rotores de hasta 8 pulgadas. La figura 3 muestra una exhibición de las esferas de acero que se podían centrifugar.

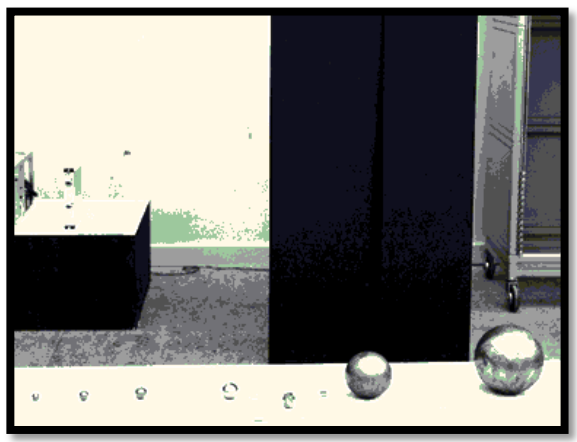

Figura 3. Fotografía de la época que muestra los diversos tamaños de los rotores.

En cuanto a la frecuencia de rotación ésta podía alcanzar 9 kilohercios para rotores de una pulgada y para rotores de 3 milímetros, centenares de kilohercios. A frecuencias mayores de 9 kilohercios los rotores de aceros más suaves experimentaban deformaciones plásticas. La figura 4 muestra uno de estos rotores después de haber sido centrifugado más allá de su límite plástico.

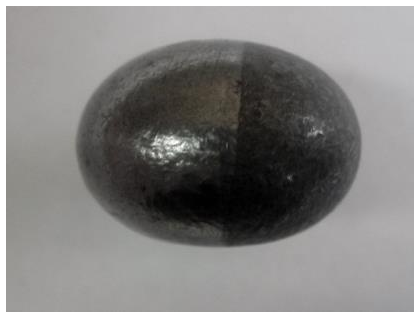

Figura 4. Rotor de acero, originalmente esférico (diámetro = 2.54 cm) deformado por centrifugación.

En algunos casos para rotores de acero duro, la fuerza centrífuga era tal que el rotor explotaba proyectando los pedazos en el blindaje de protección con el que contaba la centrífuga.

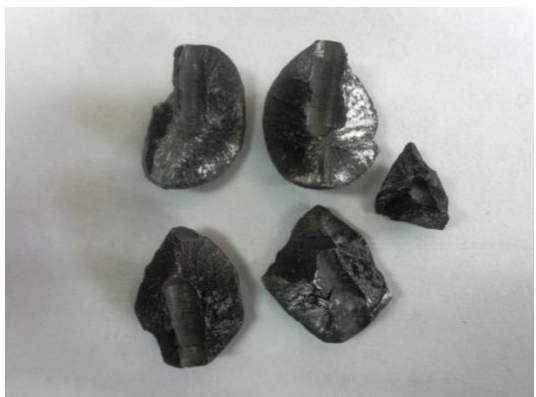
Figura 5. Fragmentos de un rotor después de explotar.

En abril de 1973 Fremerey volvió a enfocar la atención internacional hacia sus nuevos hallazgos ya que publicó en *Physical Review Letters* una carta donde decía que había construido una nueva ultracentrífuga capaz de girar una esferita de 2.5 mm cuya velocidad tangencial en su borde alcanzaba cerca de 700 m/s. Al igual que en su reporte anterior, Fremerey dejaba girar la esferita libremente a la deriva y observaba la razón de decremento Δ. En esta ocasión Fremerey reportó mediciones que parecían confirmar el efecto de frenado gravitacional predicho por Jimmy y en contraste contradecían los valores predichos por la teoría de Einstein- Eddington-Weber[63].

Sin embargo dos astrónomos, M. Reinhardt y A. Rosenblum del Instituto de Astronomía de la Universidad de Bonn (la misma universidad que la de Fremerey) se adelantaron a poner en duda la existencia del efecto de frenado predicho por Jimmy. El caso es que Fremerey mostró su manuscrito a este par de astrónomos antes de enviarlo a publicar al *Physical Review Letters*. Estos a su vez se apresuraron a enviar un artículo con sus contrargumentos al *Lettere al Nuovo Cimento*, prestigiada revista de la época. El artículo de los astrónomos apareció publicado casi dos meses antes (febrero 1973) que el de Fremerey (abril 1973)[64]. La objeción principal de Reinhardt y Rosenblum se basaba en observaciones astronómicas de la frecuencia del Pulsar de la nebulosa del Cangrejo (PSR0531+121). Un Pulsar es una estrella de neutrones que emite al girar radiación electromagnética periódica. La frecuencia de los pulsos es la medida de la frecuencia de rotación de la estrella de neutrones. De esta manera se puede considerar al Pulsar como un enorme rotor que gira rápidamente a una frecuencia observable con alta precisión. En el caso del Pulsar del Cangrejo, éste emite pulsos cada 33 milisegundos y la razón observada del decremento de su frecuencia de giro es $\Delta = 1.3 \times 10^{-11}$ s$^{-1}$. Los astrónomos hicieron la estimación de cuál sería este decremento según el modelo de Jimmy y encontraron que $\Delta = 4.69 \times 10^{-5}$ s$^{-1}$, valor muy alto comparado con el observado $(1.3 \times 10^{-11}$ s$^{-1})$[65]. Por lo tanto este par de astrónomos advierte que el valor numérico de Δ según el modelo de Jimmy, es incompatible con el valor observado, por lo que descartan que el efecto exista.

Por otro lado Eckart Frehland de la Universidad de Constanza publica un artículo (en julio de 1973) donde indica que la predicción de Jimmy se basa en un supuesto incorrecto y que el efecto simplemente no existe. [66] La suposición de Jimmy a la que se refiere Frehland dice "*la interacción de una masa puntual consigo misma debida a su aceleración es cero debido a la nula extensión del punto*". Frehland indica que esta suposición es formalmente equivalente al conocido problema de electrodinámica clásica sobre la interacción consigo misma de una carga eléctrica acelerada, que fue resuelto por Dirac en 1938 y esta interacción ciertamente no es cero. La aparición de este par de artículos debió ser un balde de agua para Fremerey y ciertamente para los propósitos de Jimmy. Incidentalmente un año después, en 1974 Russel A. Hulse y Joseph H. Taylor Jr. descubrieron un pulsar (PSR B1913+16) que presentaba anomalías en su frecuencia regular de emisión de señales. Estas anomalías fueron explicadas hacia 1978 con una increíble precisión usando la teoría de la Relatividad General de Einstein confirmando así su predicción de que una masa acelerada debía radiar energía en forma de ondas gravitacionales.[67] Este trabajo les valió a ambos el Nobel de Física 1993. Cabe señalar que el efecto aparentemente observado por Fremerey, si es que en realidad existió, quedó sin explicación aparente y su observación no se ha vuelto a repetir.

## 9 El desenlace del proyecto

Por otro lado, para poder examinar los avances que iban logrando en cuanto la separación isotópica, el grupo de ultracentrífugas realizaba periódicamente análisis de la difusión de uranio en los rotores. En términos generales, para hacer un análisis de difusión de uranio, las esferas se rebanaban paralelamente al eje de giro y la superficie de las rebanadas era analizada con la técnica de espectroscopia de masas de iones secundarios o SIMS por sus siglas en inglés, la cual proporciona información acerca de las monocapas más externas de la superficie analizada.

La elección de la técnica SIMS era la obligada ya que constituía el único método que proporciona un análisis específicamente isotópico con alta resolución y sensitividad. Además la técnica SIMS detecta todos los elementos de la tabla periódica. En términos muy simples en el SIMS, la superficie a analizar es bombardeada por un haz de iones, comúnmente gases nobles u oxígeno molecular, que al incidir sobre la superficie de la muestra producen el desprendimiento de átomos o moléculas superficiales los cuales son analizados de acuerdo con su relación carga masa (e/m) mediante un analizador de masas por deflexión magnética o cuadrupolar. El método permite analizar áreas desde unas cuantas micras hasta de milímetros cuadrados. Si el área es pequeña y la corriente del haz es considerable, la superficie se irá erosionando y de esta manera es posible analizar no sólo la superficie sino también penetrarla.

A partir de 1974 se efectuaron varias series de análisis de rotores, efectuándose las primeras en el National Bureau of Standards (NBS) mediante espectrometría de masas por ionización térmica TIMS, técnica semejante al SIMS. Los resultados obtenidos en el NBS no fueron particularmente demostrativos, sin embargo revelaron que en las esferas de aceros blandos, la difusión del uranio parecía más abundante que en la de los aceros duros. Este hecho apuntaba hacia la conveniencia del uso de rotores blandos en posteriores corridas de exploración. Por otro lado el uso de las esferas de aceros duros presentaba el serio inconveniente de que explotaban esporádicamente (ver figura 5) causando severos daños al equipo como sensores, fuentes de luz etc., mismos que se encontraban en el interior del blindaje de protección. Este tipo de eventos originaba grandes retrasos al proyecto por las reparaciones requeridas. Sin embargo Jimmy insistía en el uso de aceros duros (SKF y Marangin). La posible razón de esta insistencia es que las dimensiones físicas de las esferas de estos materiales, al ser centrifugadas, no varían en gran medida, por lo que su momento de inercia casi no varía simplificando así la posible observación del efecto gravitacional predicho por Jimmy. En contraste, los rotores de acero blando tienden a deformarse (ver figura 4). Es por esto que en 1976 el proyecto pierde a Carlos Cabrera, el electrónico de planta del proyecto, pues éste pide su cambio de adscripción a otro proyecto diferente. La razón de Cabrera es la continua negativa de Jimmy a usar rotores de aceros blandos en lugar de los duros.

Entre 1979-1980 se efectuaron análisis SIMS en el laboratorio Leybold-Heraeus en Colonia, Alemania. Básicamente estos análisis revelaron que efectivamente existía una difusión de uranio hacia la periferia pero que este se encontraba acumulado en ciertas regiones. Además los resultados no eran repetitivos. Surgió entonces la necesidad de tener un equipo SIMS propio, lo que se facilitaría el análisis de los rotores. Posteriormente, en el año de 1981, se puso en operación en el Centro Nuclear de México un equipo SIMS donde se continuaron los análisis. Con este equipo a la mano, no tardaron en darse cuenta que la difusión del uranio se efectuaba por las fronteras de grano del acero que forma la esferita del rotor. Lo que explica el por qué los resultados obtenidos por el SIMS no siempre se repetían ya que el haz de iones primarios del SIMS con el que se bombardea la muestra puede o no incidir en alguna frontera o fronteras de grano. Esto depende de la anchura del haz y de la densidad de granos de la muestra. La figura 6 hace evidente lo que ocurría en la realidad. Esta figura muestra una auto-radiografía tomada de una sección trasversal, perpendicular al eje de giro del rotor centrifugado. La auto-radiografía cubre parcialmente la barra injertada de uranio (ver figura 1) a partir de cuyo perfil circular se distingue el flujo de uranio difundido a través de las fronteras de grano (líneas zigzagueantes en la izquierda de la figura).

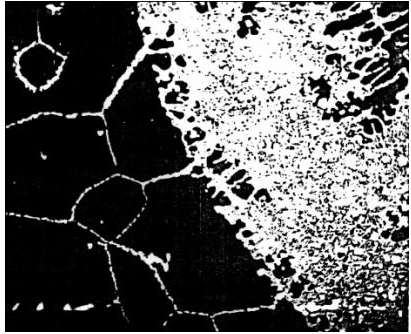

Figura 6 Auto-radiografía del corte transversal de un rotor centrifugado. A la derecha se distingue parte de la sección transversal del inserto cilíndrico de uranio (región más clara) cuyo radio es de 1.94 mm. A partir del borde circular del cilindro se aprecian las fronteras de grano (líneas blancas) por donde se ha difundido el uranio.

Esta auto-radiografía demuestra lo muy aproximadas que fueron las premisas sobre el mecanismo de difusión al arranque del proyecto. Premisas mencionadas al inicio de la sección anterior las cuales volvemos a reproducir a continuación acompañándolas con un comentario al respecto: a) *"El flujo del uranio a través de la esfera de acero resulta del movimiento átomos individuales"*. Nuestro comentario: En realidad el flujo parece ser colectivo a través de canales formados por las fronteras de grano, las cuales se enanchan al ser tensionados los granos del material por la fuerza centrífuga. b) *"Este flujo ocurre a través de un medio fijo…"*. Nuestro comentario: En realidad el medio (acero) también se mueve y se deforma (ver figura 4). c) *El flujo es cilíndricamente simétrico"*, (ver figura 2). Nuestro comentario: La realidad mostró que esta suposición es muy burda ya que la vasta mayoría de las fronteras de grano no están dispuestas a lo largo del plano de giro.

Cabe señalar que Jimmy había mostrado esporádicamente a través de informes internos, resultados prometedores. En estos informes proporcionaba datos que mostraban que en casos aislados había habido cierto grado de enriquecimiento. Sin embargo los datos presentados por él no le resultaban confiables a la Administración del Centro Nuclear en turno. El problema que el grupo de Jimmy tenía, surgía en parte del equipo SIMS con el que contaba. En pocas palabras su equipo no era confiable por tres razones: A) Había muchas interferencias moleculares de gases desprendidos de las superficies de las muestras cuyo peso molecular estaba en la vecindad del peso de los isótopos de uranio. Estas interferencias no pudieron o no supieron cómo evitarlas. B) La resolución con la que operaba el SIMS del Centro Nuclear era mayor que 1.5 dalton lo que hacía muy difícil distinguir si la cuenta en el detector provenía del uranio 235 o del 238. C) La anchura del haz de iones primarios del SIMS era bastante grande, del orden de 1 a 3 mm lo que hacía complicado examinar las acumulaciones de uranio en las fronteras de grano.

La llegada como director General del Centro Nuclear de México (ININ) en 1985 del Ing. Rubén Bello, marcó un giro en el destino y apoyo del proyecto. Como era natural, al no ser repetitivos los resultados que se obtenían en los análisis sobre contenido isotópico de las esferitas, Bello ordenó que toda prueba se sometiera a "garantía de calidad", lo cual implicaba la generación de procedimientos estrictos para cada paso de una prueba como: preparación de esferas, tratamiento térmico, encendido de la ultracentrífuga, procedimientos de operación de la ultracentrífuga, etc. Cabe señalar que este tipo de medidas administrativas es desgastante para el personal y más cuando se trata de un proyecto de investigación en donde deben explorarse métodos y procedimientos distintos a los habituales. Por otro lado, ante la sospecha de la veracidad de los datos experimentales presentados por el grupo de ultracentrífugas, la administración de Bello tomo la decisión de autorizar que una empresa especializada en análisis isotópicos hiciera los análisis correspondientes.

En 1985, la prestigiada compañía de análisis Charles Evans & Associates de California, examinó 56 muestras provenientes de 14 rotores que habían sido centrifugados por el grupo de Jimmy. Antes de ser girados, los rotores habían sido sujetos a distintos tratamientos térmicos como: el templado, recocido y el revenido, con el propósito de variar sus tamaños de grano. Las muestras analizadas consistían en "rebanadas" cortadas en paralelo al eje de giro. Las caras planas de las muestras fueron sujetas a un análisis SIMS en la compañía californiana para determinar un perfil en profundidad de la distribución isotópica del contenido de uranio de las muestras.

El equipo con que contaba la compañía Evans era fabricado por la compañía francesa CAMECA y tenía las siguientes ventajas sobre el Leybold del Centro Nuclear de México usado anteriormente para los análisis. Las ventajas del mismo eran: A) Las interferencias por moléculas que tuviesen aproximadamente el mismo peso que los isótopos del uranio, podían ser filtradas eléctrica y magnéticamente, reduciendo éstas a 1/500 con respecto al Leybold. B) La resolución del equipo era de 1/10 dalton comparada con la resolución del Leybold del Centro Nuclear que era de 1.5 dalton. C) El diámetro del haz de iones primarios del SIMS de Evans podía reducirse a tamaños entre 20 y 200 micras, lo que facilitaba examinar directamente las fronteras de grano, mientras que el del Leybold estaba entre 1 a 3 mm.

Los análisis realizados por Evans revelaron que 4 de los 14 rotores no mostraron efectos de separación isotópica, de los 8 rotores restantes sólo una pequeña parte de las rebanadas (15% de ellas) exhibieron resultados de un enriquecimiento notable de $^{235}U$. Estos efectos aparecieron en las regiones del borde del rotor y en las regiones que colindan con la barra injertada de uranio. La figura 7a muestra los resultados para el rotor marcado con la clave 4-85MB-11. Es interesante observar que a una distancia de menos de 0.25 de micra en la región adyacente a la barra insertada de uranio, el isotopo 235 presenta una abundancia de un orden de magnitud respecto al 238. Sin embargo después de esta distancia la abundancia relativa (U 235/ U 238) corresponde a la natural. Por otro lado, muestras de algunos otros rotores exhibían enriquecimiento en la periferia del rotor. La figura 7b muestra lo que ocurría en la región cercana a la superficie de la esfera para otras muestras.

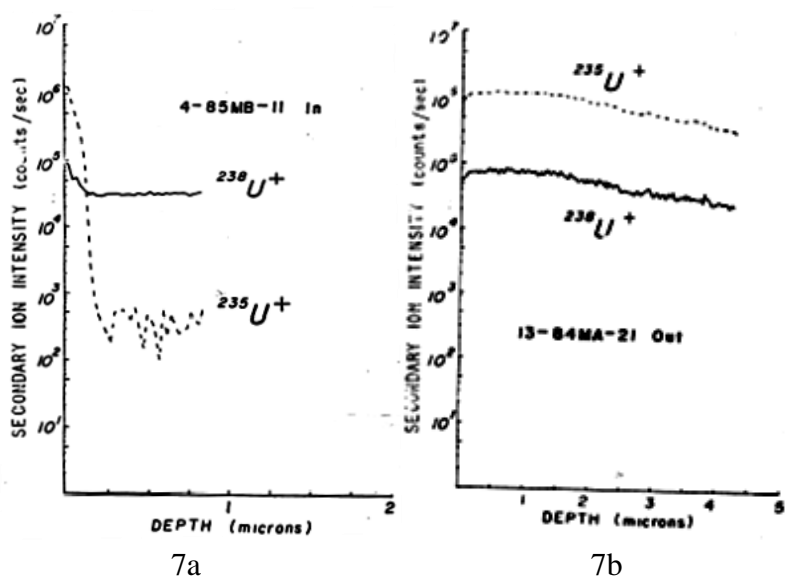

7a       7b

Figura 7. Gráficas de contenido de uranio 235 (líneas punteadas) y 238 (líneas continuas): a) en la región contigua al injerto de uranio (gráfica a la izquierda). El origen de las abscisas coincide con el límite entre la superficie de la barra injertada de uranio y el acero del rotor. El número del rotor es 4-85 MB-11. b) en la región contigua al borde del rotor (gráfica a la derecha). El origen de las abscisas coincide con el borde del rotor. El número del rotor es 13- 84 MA-21. En ambos casos la escala de las ordenadas está dada en cuentas por segundo y es directamente proporcional a la abundancia isotópica.

## 10  Comentarios Finales

En este trabajo hemos hecho referencia a dos episodios de la historia de la física del siglo XX en México: la Gravitación y el enriquecimiento de Uranio. El primero de éstos se refiere al desarrollo de una teoría alternativa a la relatividad general de Einstein propuesta por Birkhoff y desarrollada en México, principalmente por Carlos Graef-Fernández y Alberto Barajas Celis. Aunque la teoría de Birkhoff actualmente ha caído en desuso, tuvo notoriedad en su momento como ya atinadamente lo ha señalado el Dr. Octavio Obregón, reconocido experto en Gravitación y Relatividad: "[La teoría de Birkhoff] *tuvo un papel importante en la formulación y desarrollo de nuevos conceptos, así como en la confrontación con observación y experimentos*"[68].

Al respecto del comentario citado, en este trabajo hemos mostrado que las predicciones desarrolladas por Jimmy Clark Keith y fundamentadas en la teoría de Birkhoff fueron confrontados con observaciones astronómicas y con experimentos con ultracentrífugas. Sobre el primer caso, como se mencionó, diversos investigadores señalaron posibles errores en los cálculos hechos por Jimmy y en el segundo caso, la aparentemente buena confrontación entre predicción y los resultados obtenidos por Fremerey no ha sido repetida.

Por otro lado, el desarrollo de una extraordinaria ultracentrífuga transelástica en el Centro Nuclear de México, permitió demostrar que es posible enriquecer uranio con este método. Sin embargo cabe advertir que los volúmenes de las cantidades enriquecidas fueron pequeños y ocasionales. Desafortunadamente el proyecto se interrumpió cuando éste se encontraba en una etapa embrionaria y no se valoró su posible escalamiento y posterior explotación realista.

En cuanto a la producción académica del proyecto de enriquecimiento tan sólo aparecieron dos publicaciones internacionales, lo cual es entendible dado la reserva y discreción que el proyecto exigía. En ellas el grupo de Jimmy reportaba que el uranio fluía por las fronteras de grano y además señalaba algunos de los resultados que obtuvieron sobre el enriquecimiento del mismo[69,70]

En 1985, a petición del Ing. Rubén Bello (Director General del ININ en aquella época), Jimmy se traslada al Brookhaven National Laboratory, en los Estados Unidos para exponer los resultados del proyecto[71]. El 8 de agosto de 1986 Jimmy renunció a su trabajo en el Centro Nuclear de México y se integró a la Universidad de North Carolina. El proyecto en el ININ se canceló definitivamente.

---

[1] J. M. Lozano Mejía, Bol. Soc. Mex. Fis. **5**(3) (1991) 95 ,*"¿Y 50 años después?"*

[2] En la categoría de desarrollo de Instrumentos la lista de laureados fue: James Clark Keith, Roberto Jiménez Ornelas, Jaime Morales Sandoval, Ofelia Canales del Olmo y Carlos Cabrera Cruz. Recibieron una mención como colaboradores: Roberto Andrade Muñoz, Edgar Quiroz Miranda, Jaime Juárez Jiménez, Manuel Rodríguez Uribe, Humberto Rivera Meza, José D´ Oporto Zavala, Adriano Di Luca y Leopoldo Nieto Casas.

[3] *"In the field of world policy I would dedicate this Nation to the policy of the good neighbor—the neighbor who resolutely respects himself and, because he does so, respects the rights of others—the neighbor who respects his obligations and respects the sanctity of his agreements in and with a world of neighbors"*. Franklin D. Roosevelt, First Inaugural Address, 4 de marzo (1933)

[4] *"… the right of the United States to intervene in Latin America in cases of flagrant and chronic wrongdoing by a Latin American Nation"* Theodore Roosevelt, State of the Union Address, 6 de diciembre (1904)

[5] En el caso de México esto se tradujo, entre otras cosas, en el reconocimiento en 1941 por parte de los Estados Unidos, del derecho soberano de México sobre sus riquezas petroleras, terminando así el conflicto entre ambas naciones surgido por la expropiación petrolera de 1936.

[6] The Office of Coordinator of Inter-American Affairs fue establecida por Executive Order No. 8840, de Julio 30, 1941. A partir del fin de la guerra tomó el nombre de Office of Inter-American Affairs (OIAA).

[7] Manuel Sandoval Vallarta y Alfonso Nápoles Gándara fueron dos excepciones ya que fueron becados en 1927 y 1931 respectivamente. Entre los primeros físicos y matemáticos mexicanos beneficiados con las becas que la fundación John Simon Gugenhein otorgó dentro de su programa de becas a latinoamericanos podemos mencionar a: Alfredo Baños (1935,1936,1937,1957), Carlos Graef Fernández.(1937,1938,1939), Nabor Carrillo

Flores (1940,1941), Jaime LIfshitz Gaj (1942,1943), Alberto Barajas Célis (1944), Guido Münch Paniagua (1944,1945,1958), Paris Pishmish (1946), José Adem Chaim (1951, 1952), Samuel Barocio Barrios (1956)

[8] Existía y funcionaba en aquel tiempo el Observatorio Astronómico de Tacubaya cuya labor principal estaba enfocada a la astronomía y no a la astrofísica.

[9] Las partes mecánicas del telescopio fueron construidas en los talleres de la Universidad de Harvard

[10] P. Pismis "*El amanecer de la astrofísica en México*" en Historia de la Astronomía en México, varios autores (M.A. Moreno compilador) colección la Ciencia desde México F.C.E. (1986)

[11] P  "Testimonios de la Ciencia, Educación Cultura y Comunicación II La Física" , en Setenta y cinco años de Revolución, Fondo de Cultura Económica, México, **!V** (1988) p 757

[12] B. J. Bok "Astronomía Mexicana 1930-1950" en Historia de la Astronomía en México, varios autores (M.A. Moreno compilador) colección la Ciencia desde México F.C.E. (1986) ver también, B. J. Bok Rev. Mex A.A. **17** (1986) 21 "Mexican Astronomy, 1930-1950".

[13] L J Butler, "George David Birkhoff", *American National Biography* **2** (Oxford, 1999), 813-814.

[14] Carta de García a Birkhoff, Julio 20, 1939. Citada por Eduardo L. Ortiz en la Revista Saber y Tiempo **4**(15) (2003) 53, "La política interamericana de Roosevelt: George D. Birkhoff y la inclusión de América Latina en las redes matemáticas internacionales (Primera Parte)", Cf. nota 23 p.104.

[15] G.D. Birkhoff, Bol. Soc. Mat. Mex. **1** (4-5) (1944) 1.

[16] Referencia 11, p757.

[17] G.D. Birkhoff Proc. Nat. Acad. Sc. **29** (1943) 231, "Matter, Electricity and Gravitation in Flat Space-Time".

[18] *"The perfect fluid is singled out by the further requirement that the disturbance velocity is to be that of light at all densities, with the corresponding equation of state, p = ρ/2, ρ and p = f(ρ) designate, respectively, the density and pressure of the fluid"*

[19] *"Matter is supposed to be either that special, mathematically satisfactory, homogeneous adiabatic fluid for which p = ρ/2 or, presumably, any form of matter in which the disturbance velocity is that of light under all circumstances."*

[20] Ver C. M. Will Sci. Am. 231 (5)2 4(1974) "Gravitation Theory"

[21] H. Weyl, Mathematical Reviews, **4** (1943) 285.

[22] A. Barajas Proc. Nat. Acad. Sc. **30** (1944) 54 "Birkhoff´s theory of gravitation and Einstein´s theory for weak fields".

[23] H. Weyl, Proc. Nat. Acad. Sc. **30**, (1944) 205 "Comparison of a Degenerate form of Einstein´s with Birkhoff Theory of Gravitation".

[24] *"Any theory which breaks up the unity of inertia and gravitation, as Birkhoff's recent theory of gravitation in a "flat world" does, throws us back into the position before Einstein where we had to accept the identity of mass and weight without understanding it."*

[25] A. Barajas, G. D. Birkhoff, C.  Graef, M. Sandoval Vallarta, Phys. Rev. **66** (1944) 138 "*On Birkhoff´s New theory of Gravitation*".

[26] C. Graef Bol. Soc. Mat. Mex.  **1** (1944) 30 "Principios de conservación en la teoría de la gravitación de Birkhoff".

[27] G.D. Birkhoff Proc. Nat. Acad. Sc., **30** (1944) 324.

[28] M. Moshinsky Phys. Rev. **80** (1950) 514 *"Interactions of Birkhoff´s gravitational field with the electromagnetic and pair fields"*.

[29] F. Alba Andrade Rev. Mex. Fis. **1**(1) (1952) 38 "Orbitas no-planas de una partícula en el campo de una esfera de rotación en la teoría de Birkhoff".

[30] Además varios estudiantes realizaron tesis bajo la dirección de Graef a demás de otros colegas sobre campos gravitacionales de cuerpos en rotación y chocando elásticamente basados en la teoría de Birkhoff, entre ellos: Ramón Martínez Maldonado, Guillermo Aguilar Sahagún, Valentín Cardona, Fermín Viniegra, Octavio Obregón y Carlos Latorre.

[31] M. Sandoval Vallarta Bol. Soc. Mat. Mex. **1**(4 y 5) (1944) 47 "Aspectos físicos de la teoría de la gravitación de Birkhoff", reproducido en  Ciencia (México) **VI**, (1944) 9.

[32] Página 47 referencia 29

[33] "*Carlos Graef Fernández: Obra Científica*" J. L. Fernández Chapou y A. Mondragón Ballesteros, Compiladores. UAM, México (1990). Una lista aparentemente completa de los artículos publicados sobre

gravitación y relatividad en el México moderno aparece publicada en C. Ledezma Ortíz Bol. Soc. Mex. Fis **5**(3) (1991) 97

[34] *"Contra corriente: Historia de la Energía Nuclear en México (1945-1995)"* L. F. Azuela y J. L. Talancón Plaza & Valdez, México (1999)

[35] *"Los principios y las instituciones relativas al derecho de la energía nuclear. La política nuclear* A. Francoz Rigalt" p.93. México, UNAM (1988) ISBN 968-837-866-6

[36] Memoria de labores 1968-1969 p 11. Centro de documentación ININ. Archivo de información doc. 8 (1969)

[37] Keith J.C. "Gravitational Radiation in flat space-time relativity" Interim report 1g. University of Detroit. Contract No. DA-44-17-TC-519. April 30 (1959). Technical supervisión USA TRECOM (Transport Reseach Command).

[38] "[El proyecto ERG] *inspirado por el Dr. Carlos Graef Fernández y el Dr. Jesse W. Beams, llevado a cabo en la Universidad de Detroit, Michigan*." "La Ultracentrífuga Transelástica" reporte sometido a la consideración de la Academia de Investigación Científica A.C., en ocasión del Concurso Nacional de Instrumentación para la Investigación. Enero de 1974, p. 1 Introducción.

[39] Struik fue gran amigo del matemático mexicano Alfonso Nápoles Gándara, este último ex becario Gugenheim en MIT durante el período 1931. Struik realizó una visita académica a México por invitación de Nápoles Gándara en 1934 donde dictó una serie de conferencias matemáticas. La presencia de Struik en México pudo ser un factor que animó a Keith a más tarde radicar por algunos años en México y dirigir el proyecto de ultra centrifugas. Cabe también destacar que Graef Fernández y Alberto Barajas fueron discípulos de Nápoles Gándara.

[40] J. C. Keith "On Birkhoff's theory of matter and force in Minkowski space-time" B.Sc. thesis MIT, 1957.

[41] J.C. Keith Rev. Mex. Fis. **12**(1) (1963) 1 "Gravitational Radiation and Aberrated Centripetal Force Reactions in Relativity Theory: Part 2 Retarded Cohesive Forces"

[42] J. C. Keith J. Res. Nat. Bur. Stand. **67D** (1963) 533 "Magnetic torques and Coriolis effects on a magnetically suspended rotating sphere"

[43] James R Killian, Jr, "Sputnik, Scientist and Eisenhower". MIT Press (1977) Cap. 1 Sputnik and its Shock Waves.

[44] Janet Ellen Abbate, "From ARPANET to Internet: A history of ARPA-sponsored computer networks, 1966-1988" (January 1, 1994). *Dissertations available from ProQuest.* Paper AAI9503730.

[45] Cabe recordar que en aquellos años estaba en discusión el control de la gravitación. Ver por ejemplo "Guidelines to Antigravity" por R.L. Forward Am. J. Phys. **31**(3) (1963) 166. Este trabajo obtuvo el Segundo lugar en 1962, en el concurso anual organizado por la Gravity Research Foundation (http://www.gravityresearchfoundation.org/)

[46] H. A Laitinen Anal. Chem. **42**(7) (1970) 689 "Reverberations of the Mansfield Amendment".

[47] Como becario de la Comisión Nacional de Energía Nuclear de México (beca 504.111/198)

[48] J.C. Clark. Reporte especial para la Coordinación General del INEN, "Proyectos realizados, actuales y futuros" 2 mayo 1977.

[49] F.A. Lindemann and F.W. Aston, Phil. Mag. **37**(6) (1919) 523.

[50] J.W. Beams and F.B. Haynes Phys. Rev. **50** (1936) 491.

[51] R. Scott Kemp "Gas centrifuge theory and development: a review of U. S. programs" en Science and Security **17**(2009) 1. *DOI: 10.1080/08929880802335816*

[52] Norman Gall "Atoms for Brazil, dangers for all" Bull. Atom. Scien. **32**(6) (1976) 4.

[53] F.T. Holmes, Rev. Sci. Instr. **8** (1937) 444.

[54] F.T. Holmes and J.W. Beams Nature **140** (1937) 30.

[55] Por ejemplo Beams reportó que "*un rotor girando rápidamente perderá varias revoluciones por segundo si una puerta es azotada en un cuarto contiguo. "…a rapidly spinning rotor will lose several rps if a door is slammed in the next room*". J.W. Beams, D.M.Spitzer Jr., and J.P. Wade Jr., Rev. Sci. Instrum. **33** (1962) 151.

[56] Como ya lo mencionamos, el programa ERG de la Universidad de Detroit según testimonio del propio Jimmy, fue inspirado por Carlos Graef y Jesse W. Beams. Ref. 38.

[57] Ver por ejemplo el trabajo Isotope Separation by Condensed Matter Centrifugation: Sedimentation of Isotope Atoms in Se" de T. Mashimo, M. Ono, X. Huang, Y. Iguchi, S. Okayasu, K. Kobayashi and E. Nakamura "J. Nuc. Sc. and Tech., Suppl. **6** (2008) 105.

[58] J.K. Fremerey Rev. Sci. Inst. **42**(6) (1971) 753, "Apparatus for Determination of Residual Drag Torque on Small Spinning Spheres in Magnetic Suspension".

[59] Fremerey trabajó desde 1963 -1974 en el laboratorio de ultracentrífugas de gas con Wilhelm Groth, personaje al cual ya nos referimos en relación a la venta de estos instrumentos al Brasil.

[60] "*The reliability must be increased up to the weekend stability in order to be able to look for a gravitational effect, which should appear according to Keith´s theoretical considerations*". Fremerey (1971) op. cit. ref 53, p.762.

[61] "La Ultracentrífuga Sólida y su Aplicación en la Separación de Isótopos de Uranio: Memorias" 10 de junio de 1982 ININ.

[62] Ver nota 2.

[63] J.K. Fremerey, Phys. Rev. Lett. **30** (1973) 753, " Significant Deviation of Rotational Decay from Theory at a Reliability in the 10-12 sec-1 Range"

[64] M Reinhardt and A. Rosenblum Lett. Nuovo Cimento **6**(5) (1973) 189, "The nonexistence of a Relativistic effect proposed by Keith".

[65] Cf. Ecuación 1, $\Delta = -\frac{1}{2}k(a\omega/c)^3\omega = 4.69 \times 10^{-5}$ s$^{-1}$, donde k = 75π/128 para un rotor esférico, *a* =10 km el radio de giro del pulsar, *T*= (2πω)$^{-1}$ = 33 ms el período observado de giro del pulsar y c = 3×10$^5$ km s$^{-1}$ la velocidad de la luz.

[66] E. Frehland Lett. Nuovo Cimento **7**(12) (1973) 490, "Critique of the Gravitational Radiation Damping Effects Calculated by Keith".

[67] Para una explicación detallada del fenómeno ver por ejemplo, J.M. Wiesberg, J. H. Taylor and L. A. Fowler Sci. Am. **245**(4) (1981) 66 "Gravitational Waves from an Orbiting Pulsar".

[68] O. J. Obregón Díaz, "El nacimiento de la Relatividad y la Gravitación en México" en "La relatividad en México Ma. P. Ramos Lara ed. (2008) UNAM ISBN 978-970-32-4608-3

[69] J.C. Keith, M. I. Mora, O. Olea and J.L. Peña Springer Series in Chemical Physics **44** (1986) 409. "SIMS trace detection of heavy elements in high-speed rotors".

[70] J.C. Keith, O. Olea and J.L. Peña Proceedings of the 10th International Vacuum Congress (IVC-10), 6th International Conference on Solid Surfaces (ICSS-6) and 33rd National Symposium of the American Vacuum Society. October 27-31,1986. Baltimore, Maryland USA ISBN 0734-2101

[71] O. Olea. "Difusión forzada de elementos pesados en sólidos girados a altas velocidades soportados magnéticamente" Reporte Interno, invierno 1986. Centro de Información y Documentación Nuclear (ININ) Documento No. Olea 5557_1624.pdf, p12.